\documentclass[aps,prb,twocolumn,showpacs,amsmath,amssymb,footinbib,longbibliography]{revtex4-2}

\usepackage[percent]{overpic} 
\usepackage{graphicx}

\usepackage{bm}         
\usepackage[dvipsnames]{xcolor} 
\usepackage{tikz}       
\usepackage{mathrsfs}
\usepackage{subfigure, psfrag}
\usepackage[colorlinks=true,urlcolor=blue,citecolor=blue,linkcolor=blue]{hyperref}
\hypersetup{breaklinks=true}

\begin{document}
\title {Longitudinal magnon transport properties in the easy-axis XXZ Heisenberg ferromagnet on the face-centered cubic lattice}
\author{M.~R.~Parymuda}
\email{mparymuda@icmp.lviv.ua\\
}
\affiliation{Institute for Condensed Matter Physics, National Academy of Sciences of Ukraine, Svientsitskii Street 1, 79011, Lviv, Ukraine}
\date{\today}
\begin{abstract}
We present a detailed investigation of longitudinal magneto-thermal transport in the $S=1/2$ ferromagnetic XXZ model with easy-axis exchange anisotropy ($\Delta>1$) on a face-centered cubic lattice consisting of four sublattices. We employ linear spin-wave theory and the Kubo formalism to evaluate the longitudinal spin and thermal conductivities, both of which exhibit activated temperature dependence in the low-temperature regime, and to determine their magnetic-field dependence.   
Our analysis indicates that a magnon gap is crucial for ensuring the convergence of these conductivities. 
Furthermore, by examining the ratio of thermal conductivity to spin conductivity, we identify an analog of the Wiedemann-Franz law for magnon transport at low temperatures. Finally, we demonstrate that these results can be generalized to systems with arbitrary spin.
\end{abstract}
\maketitle
\section{Introduction}
The study of spin-wave transport in quantum materials remains one of the active areas of condensed matter physics, driven by both fundamental interest and potential technological applications in magnon spintronics \cite{kruglyak2010,chumak2015,jungwirth2016,chumak2017,flebus2024}. In particular, exploring how spin excitations propagate in magnetically ordered systems provides valuable insights into the interplay between spin dynamics and lattice structure.

Magnetic insulators offer significant advantages over metals for transmitting spin currents via spin waves (magnons). 
The absence of free charge carriers in insulators reduces energy dissipation, enabling spin waves to propagate over long distances with reduced Gilbert damping parameter. This inherent efficiency makes magnetic insulators particularly attractive for applications in magnonics. Prominent examples of low-damping materials are yttrium iron garnet Y$_3$Fe$_5$O$_{12}$(YIG) \cite{serga2010,cornelissen2015,cherepanov1993},  europium oxide EuO \cite{makino2012}, europium sulfide EuS \cite{Anguilar-Pujol2023}, manganese ferrite MnFe$_{2}$O$_{4}$ \cite{hong2024}, chromia Cr$_{2}$O$_{3}$ \cite{belashchenko2016}, and copper oxide selenite Cu$_2$OSeO$_3$ \cite{stasinopoulos2017}.

Linear spin-wave theory (LSWT) provides a powerful analytical framework for investigating the low-temperature behavior of magnetic systems. Within LSWT \cite{holstein1940,aoya2019,aoya2022,aoki2004,sentef2007,rez}, the system's spin operators are mapped onto bosonic creation and annihilation operators via Holstein-Primakoff (HP) transformation, applied to leading order. This reformulation treats magnons as non-interacting bosons, significantly simplifying problem. The approximation is valid specifically at low temperatures, where magnon density remains extremely small.

\begin{figure}[t]
\includegraphics[width=1.1\linewidth]{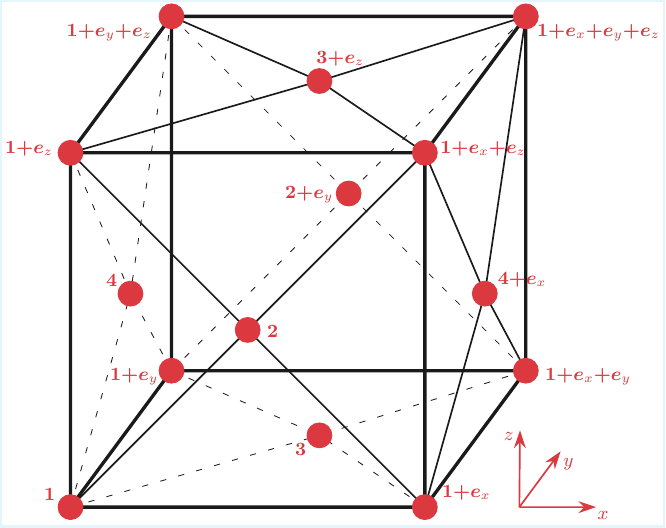}

\caption{Schematic representation of our FCC lattice with four sublattices per unit cell. Nearest-neighbor (NN) bonds are shown by thin and dashed lines, while the next-nearest-neighbor (NNN) bonds are indicated by solid lines.}
\label{fig:Fig1}
\end{figure}

Many rock-salt compounds host magnetic ions on the face-centered cubic (FCC) lattice. In particular, rare-earth chalcogenides such as EuO \cite{makino2012} and EuS \cite{Anguilar-Pujol2023} are FCC ferromagnets with a low Gilbert damping parameter. In these materials, exchange anisotropy can arise from spin-orbit interaction, making the XXZ model a suitable first approximation \cite{skomski,Lado}. The FCC lattice therefore serves as a starting point for studying the low-temperature properties of the XXZ Heisenberg ferromagnet.

We study the XXZ model on the three-dimensional FCC lattice \cite{ackland2023,hattori2023} (see Fig.~\ref{fig:Fig1}), where spins are located at both the cube vertices and the centers of its faces. This geometry results in a cubic unit cell composed of four sublattices that give rise to both intra- and inter-sublattice exchange interactions. The FCC lattice thus provides an ideal platform for investigating how the interplay between anisotropic spin interaction and lattice geometry affects the system's transport properties. Using a combination of methods, namely LSWT and the Kubo formalism, we compute the temperature dependence of both the spin and thermal conductivity in the low-temperature regime. In the zero-field case, these conductivities exhibit thermally activated behavior. Furthermore, we derive a magnon analog of the Wiedemann-Franz (WF) law \cite{nakata2015,nakata2017,nakata2017(1),nakata2022}, which links the thermal conductivity $\kappa$ and spin conductivity $\sigma$ through the universal proportionality constant known as the Lorentz number. We also demonstrate the magnetic-field dependence of the spin and thermal conductivity.

Magnon transport properties and thermodynamic properties in a ferromagnet are coupled through the influence of an external magnetic field. The field tunes a magnon gap and a magnon density, which in turn govern both the entropy and the transport properties. This interplay provides a microscopic mechanism for the magnetocaloric effect, as the field-induced suppression of the magnon density reduces the entropy and transport conductivities of the system at fixed temperature $T$ \cite{pecharsky,shi2024,shi2024(1),shi2025}.

The remainder of this paper is organized as follows. In Section II, we introduce the ferromagnetic XXZ Hamiltonian that serves as the foundation for our study. We explicitly formulate this anisotropic spin model, detailing its exchange interactions between spins within the unit cell depicted in Fig.~\ref{fig:Fig1}. In Section III, we derive the magnon Hamiltonian by applying LSWT. 
In Section IV, we present a comprehensive analysis of temperature and magnetic field dependence of the spin and the thermal conductivities in the low-temperature limit. In Section V, we conclude by summarizing our main findings and comparing them with the results in the literature.

In what follows, we set $\hbar =1$.
\section{Model}
The positions of the lattice sites are defined by $\bm R_{\bm m\alpha}=\bm R_{\bm m}+\bm r_{\alpha}$. Here, the unit cell positions are given by $\bm R_{\bm m}=m_x \bm e_x+m_y \bm e_y+m_z \bm e_z$, where $m_x, m_y, m_z$ are integers that define the cells along the three directions. The unit vectors are defined as $\bm e_x=(1,0,0)$, $\bm e_y=(0,1,0)$, and $\bm e_z=(0,0,1)$. The displacement vectors $\bm r_\alpha$, where $\alpha$ is sublattice number, specify the relative positions of the sublattice sites within each unit cell. The lattice constant is set to unity ($a=1$).

The exchange Hamiltonian \cite{ackland2023} is given by
\begin{equation}
    \mathcal H=J_1\sum_{\langle \bm m\alpha;\bm n\beta\rangle} \bm S_{\bm m\alpha} \cdot \bm S_{\bm n\beta}+J_2\sum_{\langle\langle \bm m\alpha;\bm n\beta\rangle\rangle}\bm S_{\bm m\alpha}\cdot \bm S_{\bm n\beta},
    \label{eq:1}
\end{equation}
where all exchange interaction bonds are ferromagnetic i.e., $J_1<0$ and $J_2<0$. The first sum in Hamiltonian (\ref{eq:1}) runs over all NN pairs, while the second sum runs over all NNN pairs.

The anisotropic scalar product is defined as follows 
\begin{equation}
\bm S_{\bm m\alpha}\cdot \bm S_{\bm n\beta}=\frac{1}{2}(S^+_{\bm m\alpha}S^-_{\bm n\beta}+S^-_{\bm m\alpha}S^+_{\bm n\beta})+\Delta S_{\bm m\alpha}^zS_{\bm n\beta}^z,
\end{equation}
where $S^\pm_{\bm m\alpha}=S^x_{\bm m\alpha}\pm iS^y_{\bm m\alpha}$, and the anisotropy parameter $\Delta=J_z/J$, with $J_x=J_y=J$. For simplicity, we set $\Delta_1=\Delta_2=\Delta$.

Applying Hamiltonian~(\ref{eq:1}) to the FCC lattice (Fig.~\ref{fig:Fig1} ) in the presence of an external magnetic field, we obtain the following Hamiltonian
\begin{widetext}
\begin{equation*}
\mathcal H=-|J_1|\sum_{\bm m}\Big[(\bm S_{\bm m1}+\bm S_{\bm m+\bm e_x,1}+\bm S_{\bm m+\bm e_z,1}+\bm S_{\bm m+\bm e_x+\bm e_z ,1} )\cdot \bm S_{\bm m2}+(\bm S_{\bm m1}+\bm S_{\bm m+\bm e_x,1}+\bm S_{\bm m+\bm e_y,1}+\bm S_{\bm m+\bm e_x+\bm e_y,1})\cdot \bm S_{\bm m3}
\end{equation*}
\begin{equation}
 +(\bm S_{\bm m1}+\bm S_{\bm m+\bm e_y,1}+\bm S_{\bm m+\bm e_z,1}+\bm S_{\bm m+\bm e_y+\bm e_z,1})\cdot \bm S_{\bm m4}\Big]-|J_2|\sum_{\bm m}(\bm S_{\bm m+\bm e_x,1}+\bm S_{\bm m+\bm e_y,1}+\bm S_{\bm m+\bm e_z,1})\cdot \bm S_{\bm m1}-g\mu_B H\sum_{\bm m}\sum_{\alpha=1}^{4}S^z_{\bm m\alpha}.
 \label{eq:2}
\end{equation}
\end{widetext}
We take the magnetic field $H$ along the $z$ axis. Here, $g$ is the Lande factor and $\mu_B$ is the Bohr magneton.
\section{Magnon Hamiltonian}
Applying the HP transformation to leading order, the spin operators become 
\begin{equation}
 S_{\bm m\alpha}^+=\sqrt{2S}a_{\bm m\alpha},\;\;\; S_{\bm m\alpha}^-=\sqrt{2S}a^\dagger_{\bm m\alpha} ,\;\;\; S_{\bm m\alpha}^z=S-a^\dagger_{\bm m\alpha}a_{\bm m\alpha}.   
\end{equation}
The introduced bosonic operators satisfy the following commutation relation 
\begin{equation}
    [a_{\bm m\alpha},a^\dagger_{\bm n\beta}]=\delta_{\bm m\bm n}\delta_{\alpha\beta}.
\end{equation}

We introduce the Fourier transform of the bosonic operators 
\begin{equation}
    a_{\bm m\alpha}=\frac{1}{\sqrt{\mathcal N}}\sum_{\bm q}e^{i \bm q\cdot \bm R_{\bm m}}a_{\bm q\alpha},
\end{equation}
where $\mathcal N=N/4$ is the number of unit cells, $N$ is the total number of spins, and the momentum components are given by $q_\gamma=2\pi n_\gamma/\mathcal L_\gamma$, for $\gamma=x, y, z$ with $n_\gamma=1,\hdots,\mathcal L_\gamma$. We note that $\mathcal N=\mathcal L_x \mathcal L_y \mathcal L_z$.

\begin{figure}[]
  \centering
  \subfigure[]{\includegraphics[width=0.45\textwidth]{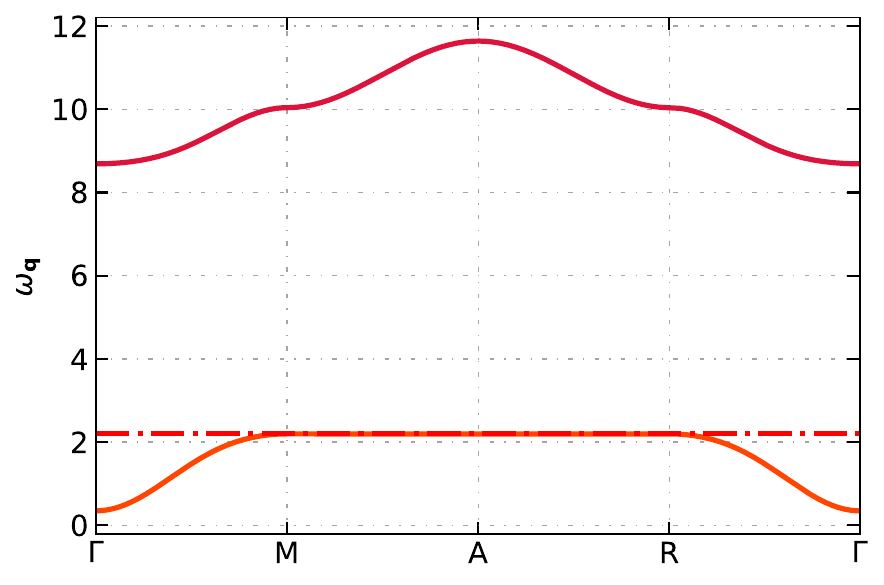}}
  \subfigure[]{\includegraphics[width=0.39\textwidth]{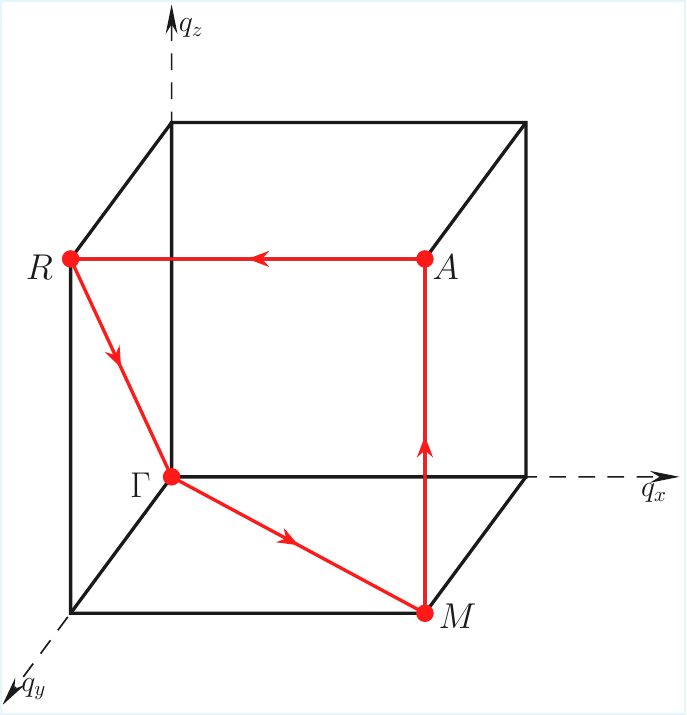}}
 \caption{(a) Magnon dispersion relations along the high-symmetry path $\Gamma-M-A-R-\Gamma$ in the first Brillouin zone. The symmetry points are $\Gamma=(0,0,0)$, $M=(\pi,\pi,0)$, $A=(\pi,\pi,\pi)$, and $R=(0,\pi,\pi)$. Here, $\Delta=1.1$, $|J_1|=1$, $|J_2|=0.8$, and $H=0$. The upper and lower branches are $\omega_{\bm q+}$ and $\omega_{\bm q-}$, respectively. (b) Schematic representation of the high-symmetry path in the first octant of the Brillouin zone ($q_x, q_y, q_z>0$).}
  \label{fig:Fig2}
  \end{figure}

In momentum space, the exchange interaction between spins on sublattice $\alpha$ and $\beta$ separated by the vector  $\bm e_\gamma$ can be written as 
\begin{equation*}
\sum_{\bm m}{\bm S}_{{\bm m+\bm e_\gamma},\alpha}\cdot {\bm S}_{{\bm m}\beta}=\mathcal N S^2\Delta+S\sum_{\bm q}(e^{iq_\gamma} a_{\bm q \alpha}a^\dagger_{\bm q\beta}+e^{-iq_\gamma} a^\dagger_{\bm q \alpha}a_{\bm q\beta}
\end{equation*}
\begin{equation}
-\Delta a^\dagger_{\bm q\alpha}a_{\bm q\alpha}-\Delta a^\dagger_{\bm q\beta}a_{\bm q\beta}).
\label{eq6}
\end{equation}
Repeating this procedure for all 15 bonds, we can write our Hamiltonian in a more compact form  as follows 
\begin{equation}
\mathcal H=\mathcal H_0+ \sum_{\bm q}\begin{pmatrix}
a_{\bm q1}^\dagger &a^\dagger_{{\bm q}2}& a_{\bm q3}^\dagger & a^\dagger_{{\bm q}4} 
\end{pmatrix}
 \bm{M}
 \begin{pmatrix}
a_{\bm q1}\\\\
a_{\bm q2}\\\\
a_{\bm q3}\\\\
a_{{\bm q}4}

\end{pmatrix},
\label{eq7}
\end{equation}

where
\begin{widetext}

\begin{equation}\bm {M} =-
\begin{pmatrix}
|J_2|\left(\cos q_x+\cos q_y+\cos q_z\right)-3\left(|J_2|+2|J_1|\right)\Delta-g\mu_B H& \phi^*_{xz} & \phi^*_{xy}& \phi^*_{yz}  \\
\phi_{xz}  & -2|J_1|\Delta-g\mu_B H & 0 &0\\
\phi_{xy} & 0 & -2|J_1|\Delta-g\mu_B H &0 \\
\phi_{yz} &0&0&-2|J_1|\Delta-g\mu_B H
\end{pmatrix},
\label{eq8}
\end{equation}
and 
\begin{equation}
\phi_{\alpha\beta}=|J_1|(1+e^{iq_\alpha}+e^{iq_\beta}+e^{i(q_\alpha+q_\beta)})/2.
\label{eq9}
\end{equation}

By finding the eigenvalues of the matrix $\bm M$, we obtain the following dispersion relations

\begin{equation}\omega_{\bm q1}=\omega_{\bm q2}=2|J_1|\Delta+g\mu_BH,
\label{eq14}
\end{equation}
\begin{equation}
\omega_{\bm q\pm}=\frac{|J_2|}{2}\left(\left(3+8\delta\right)\Delta-\cos q_x-\cos q_y-\cos q_z\pm S\right)+g\mu_B H,
\label{eq15}
\end{equation}
where 
\begin{equation}
S=\sqrt{\left(\cos q_x+\cos q_y+\cos q_z+4\delta^2-\Delta(3+4\delta)\right)^2+4\delta^2\left(3-4\delta^2+2\Delta  (3+4\delta)+\cos q_x\cos q_y+\cos q_x \cos q_z+\cos q_y \cos q_z\right)},
\end{equation}
\begin{equation}
\delta=|J_1|/|J_2|.
\end{equation}
\end{widetext}
Throughout the paper, we set $|J_1|=|J|$ and $|J_2|=0.8|J|$, where $|J|$ is the energy scale. The magnon dispersion relations $\omega_{\bm q}$ are shown as functions of momentum in Fig.~\ref{fig:Fig2}(a) along the high-symmetry path in the first Brillouin zone (see Fig.~\ref{fig:Fig2}(b)).

We restrict our analysis to low-energy (low-momentum) excitations, as at low temperatures thermally excited magnons predominantly populate states near the magnon band minima \cite{kittel, barker}.

By finding the eigenvalues of the matrix in the asymptotic long-wavelength limit ($|\bm q|\to 0$), we obtain the following spin-wave (magnon) dispersion relations 
\begin{equation}
\omega_{{\bm q}\pm}=\Delta^\pm_{gp}+D^\pm|{\bm q}|^2 ,
\label{eq11}
\end{equation}
where the explicit forms of the gaps $\Delta^\pm_{gp}$ and stiffness constants $D^\pm$ are given in Appendix~\ref{sec:A}. To restore the lattice constant, one needs to replace $D^\pm\to a^2 D^\pm$. 

The first two branches are dispersionless, indicating that these modes are localized and are determined solely by the exchange interaction, the anisotropy parameter $\Delta$, the $g$-factor, and the external magnetic field, while the latter two branches exhibit a quadratic momentum dependence. The presence of the finite gap in these dispersive branches is crucial for evaluating transport coefficients, as will be shown in subsequent sections.

The above dispersion relations demonstrate that the anisotropy parameter $\Delta$ significantly modifies the expressions for the spin-wave (magnon) gaps. In fact, for $\Delta=1$, the Hamiltonian~\eqref{eq:2} reduces to the isotropic Heisenberg model, whose dispersion relations comprise two flat (dispersionless) branches and two dispersive branches, one of which remains gapless (acoustic mode). In the Heisenberg model, the external magnetic field is one of the mechanisms capable of opening the gap in both dispersive branches of the system. By contrast, for $\Delta>1$, both dispersive branches acquire finite gaps even in the absence of any applied field.

To proceed, we transform to the diagonal basis using a unitary matrix $\bm U$ composed of the eigenvectors of the matrix $\bm M$. In fact, we can bring the matrix $\bm M$ into diagonal form 
\begin{equation}
    \bm U^\dagger \bm M \bm U= diag(\omega_{\bm q1},\hdots, \omega_{\bm q4}).
\end{equation}

In the diagonal basis, the Hamiltonian is expressed in terms of the new magnon operators $\bm b^\dagger=\bm a^\dagger \bm U$ and $\bm b=\bm U^\dagger\bm a $ as follows
\begin{equation}
    \mathcal H=\mathcal H_0+\sum_{\bm q}\sum_{\alpha=1}^{4}\omega_{\bm q\alpha}b^\dagger_{\bm q\alpha}b_{\bm q\alpha},
    \label{eq:14}
\end{equation}
where 
\begin{equation}
    \mathcal H_0=-\frac{\mathcal N \Delta}{4}(12|J_1|+3|J_2|)-4\mu_B\mathcal N H.
    \label{eq:15}
\end{equation}
In Hamiltonian~(\ref{eq:14}), the operators $b_{\bm q\alpha}$ and $b^\dagger_{\bm q\alpha}$ annihilate and create a magnon with momentum $\bm q$ on sublattice $\alpha$, respectively. The term $\mathcal H_0$ contributes only to the ground-state energy and does not affect the system's transport properties.


\section{Magnon transport properties}
\subsection{Zero magnetic field}
In the low-temperature limit ($\Delta_{gp}/k_BT \to \infty$), the spin conductivity exhibits thermally activated behavior described by the asymptotic law $\sigma(T)\propto T^{3/2}e^{-\Delta_{gp}/k_BT}$ (see Appendix~\ref{subsec:B1}), as demonstrated in Fig.\ref{fig:Fig3}(a) for various values of the anisotropy parameter. This dependence reveals high sensitivity of the spin conductivity to the absence of the gap in the magnon dispersion relation and indicates that exponential behavior dominates at low temperatures. In fact, for the Heisenberg model ($\Delta=1$), one of the dispersive branches is gapless, resulting in a divergence of the spin conductivity $\sigma \to \infty$.
\begin{figure*}[t]
  \centering
  \subfigure[]{\includegraphics[width=0.47\textwidth]{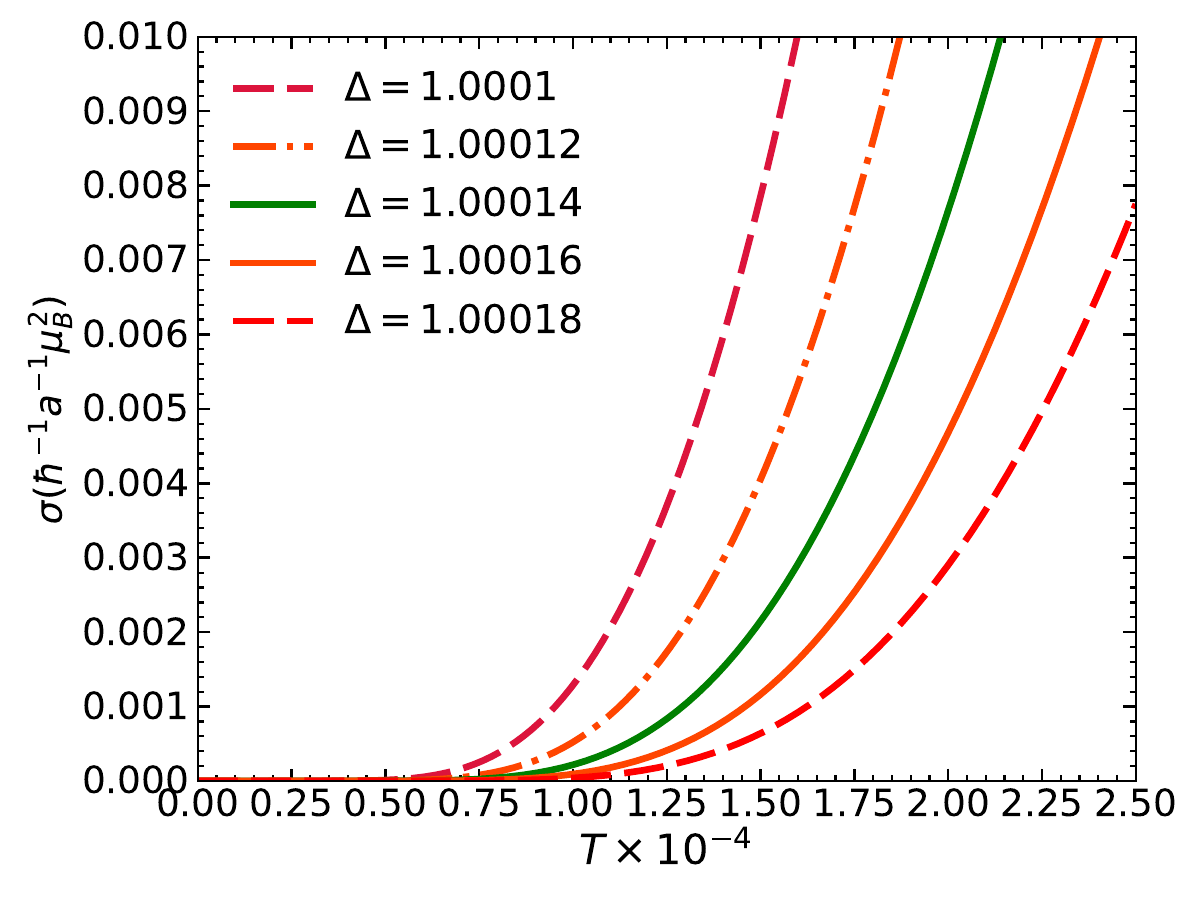}}
  \subfigure[]{\includegraphics[width=0.47\textwidth]{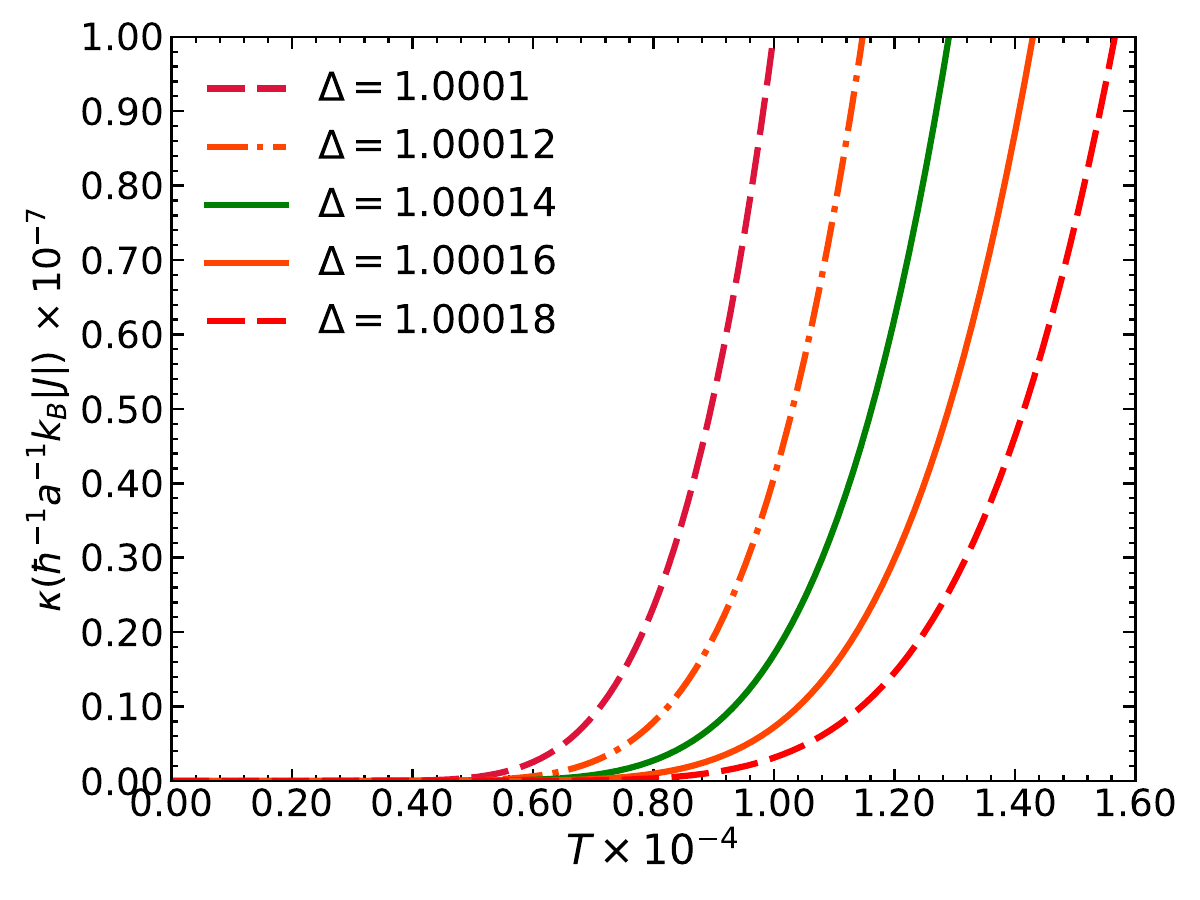}}
  \caption{The temperature dependence of (a) spin conductivity $\sigma$ and (b) thermal conductivity $\kappa$ are shown for several values of the anisotropy parameter $\Delta$. The temperature is measured in dimensionless units $T \equiv k_B T/|J|$. The parameters used are $\alpha=0.01$, $|J_1|=|J|$, $|J_2|=0.8|J|$, and $g=2$ ( $S=1/2$). }
  \label{fig:Fig3}
  \end{figure*}

 The thermal conductivity of the system exhibits similarly activated behavior at low temperatures ($\Delta_{gp}/k_BT \to \infty$), obeying the asymptotic law $\kappa(T) \propto T^{5/2}e^{-\Delta_{gp}/k_BT}$ (see Appendix~\ref{subsec:B1}), as demonstrated in Fig.\ref{fig:Fig3}(b). Like the spin conductivity, the thermal conductivity is highly sensitive to the magnon gap in the magnon dispersion expressions. Consequently, for the Heisenberg model ($\Delta=1$), where the gap vanishes, thermal conductivity shows an analogous divergence $\kappa \to \infty$. 

At $T=0$, both the spin conductivity $\sigma(T)$ and thermal conductivity $\kappa(T)$ vanish due to the absence of thermally excited magnons. In the low-temperature regime ($k_BT\ll\Delta_{gp}$), the Bose distribution function for each magnon branch can be approximated as $n_{B}\simeq e^{-\omega_{\bm q\alpha}/k_BT}$. For optical branches ($\Delta_{gp}\neq 0$), the magnon density is suppressed by the presence of a gap in the magnon dispersion relation and scales as $N_{\rm mag} \propto T^{3/2}e^{-\Delta_{gp}/k_BT}$. By contrast, the magnon density for acoustic branches ($\Delta_{gp}=0$) follows a power-law behavior $N_{\rm mag} \propto T^{3/2}$. Although the magnon density of the acoustic (gapless) mode is significantly larger than that of the optical (gapped) mode and is finite, the corresponding spin and thermal conductivities for gappless branches diverge. Moreover, while flat bands have a very high density of states, their group velocity vanishes everywhere on the flat band, and thus they do not contribute to spin or thermal conductivities. 

However, when $k_BT \sim \Delta_{gp}$, the activated behavior breaks down. The magnon population becomes finite and interactions between magnons must be taken into account.

At low temperatures, the lower branch dominates (see Fig.\ref{fig:Fig2}) due to thermally activated behavior and because the gap for the upper branch $\omega_{\bm q+}$ is much larger  than that for the lower branch $\omega_{\bm q-}$ (i.e.,  $\Delta_{gp}^+\gg \Delta_{gp}^-$). As a result, the spin conductivity $\sigma(T)$ and thermal conductivity $\kappa(T)$ are entirely determined by the contribution from the lower branch since the magnon occupation number of the upper branch is exponentially reduced. Although a gap is necessary to ensure the convergence of these conductivities, optimal spin and thermal conductivities require
this gap to be very small (i.e., $\Delta_{gp} \ll1$).

In our analysis, we consider a topologically trivial ferromagnet (with preserved inversion symmetry) on the three-dimensional FCC lattice. This implies that the transverse components of the spin conductivity $\sigma(T)$ and thermal conductivity $\kappa(T)$ vanish. In contrast, previous studies have examined the two-dimensional checkerboard lattice \cite{pires2021, zhu2010}, in which inversion symmetry is broken, thus allowing transverse transport to emerge.

It is also worth noting that the longitudinal spin and thermal conductivities diverge in two-dimensional systems, even when Gilbert damping parameter $\alpha \ll 1$ and a gap is present in the system. This divergence arises from the infrared singularity of the Bose function at zero frequency, which yields an infinite contribution to these conductivities.

Considering the derived temperature dependencies of the spin and thermal conductivity, we obtain the relation $L=\kappa/\sigma T$, where $L=5/2$ in units of $(k_B/g\mu_B)^2$ is a Lorentz number that captures the universal features of magnon transport in the low-temperature regime. The detailed derivation of $L$ is given in Appendix~\ref{subsec:B1}. This confirms that the anisotropy parameter $\Delta$ does not affect the WF law.

\begin{figure*}[t]
  \centering
  \subfigure[]{\includegraphics[width=0.47\textwidth]{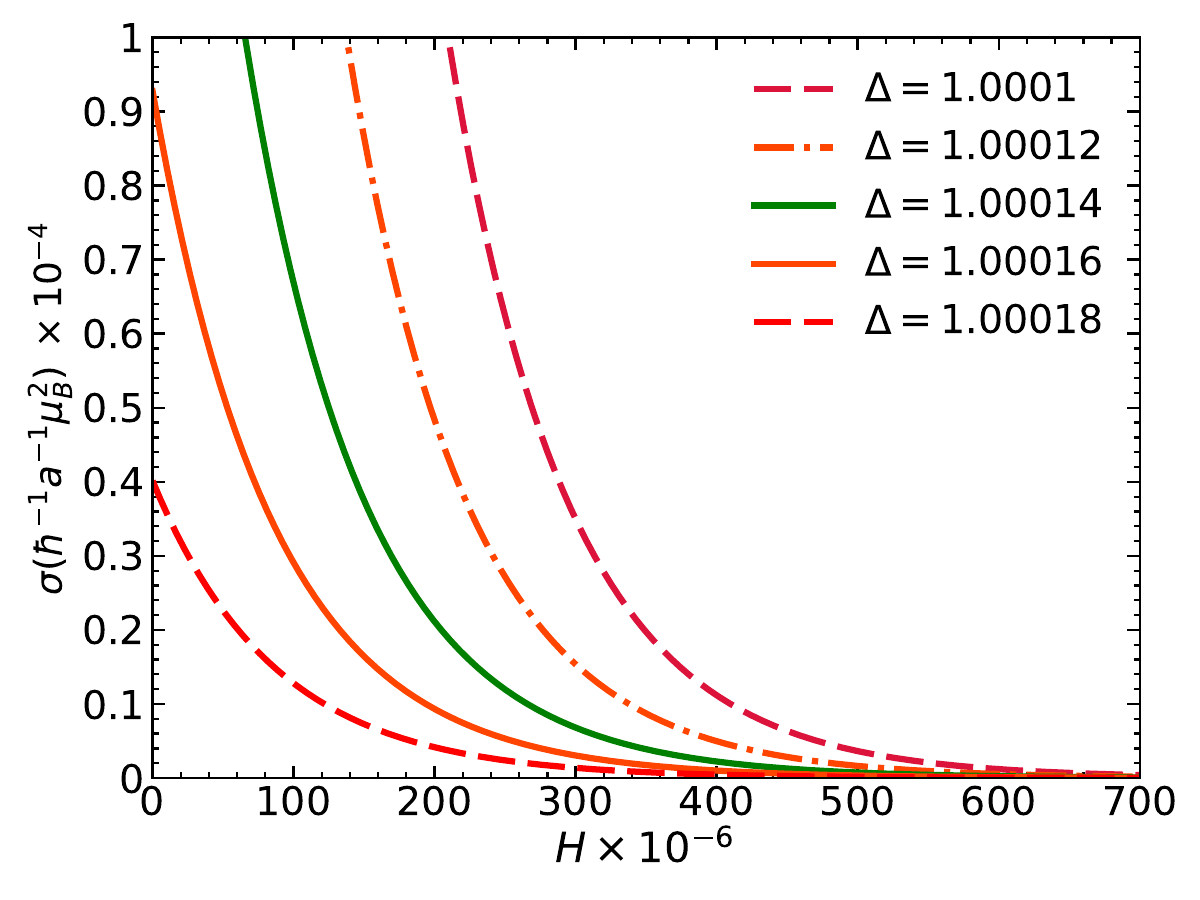}}
  \subfigure[]{\includegraphics[width=0.47\textwidth]{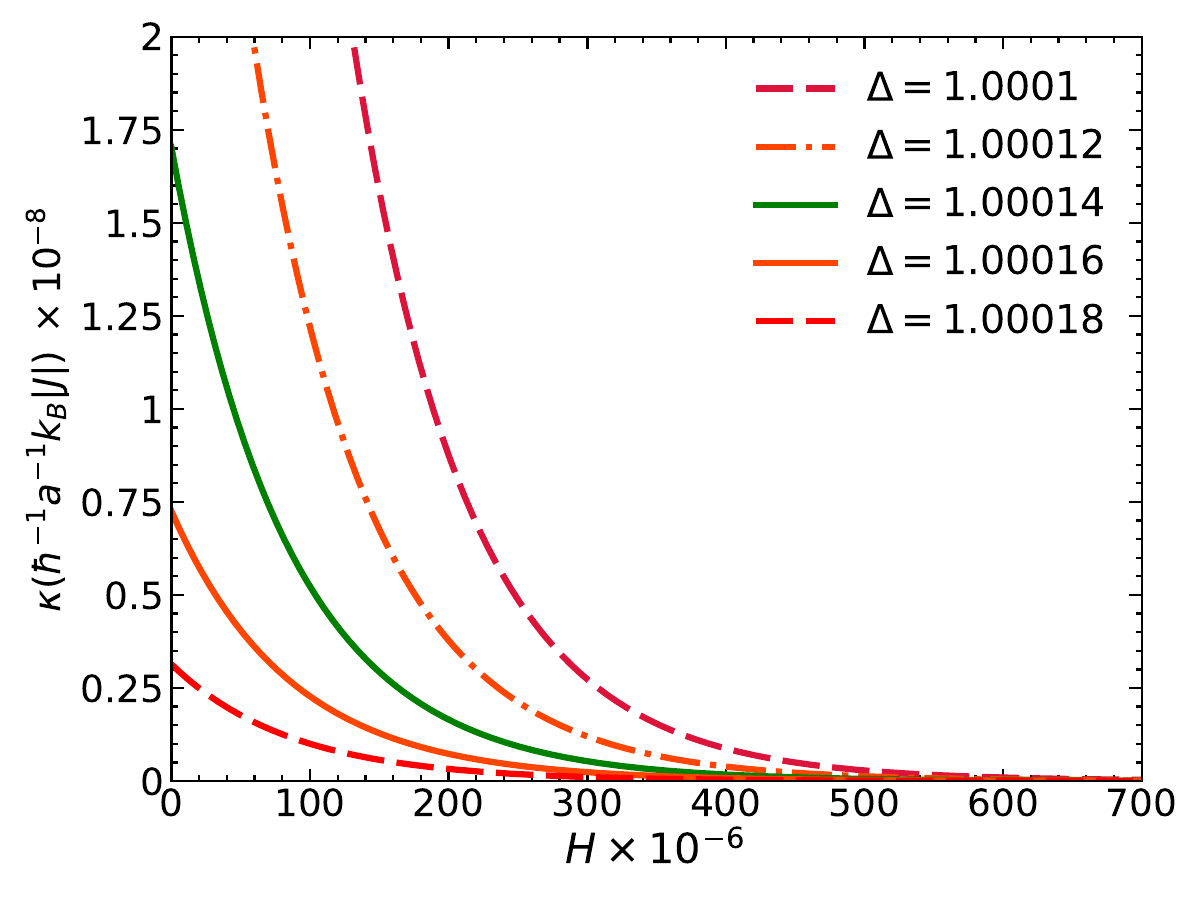}}
  \caption{The magnetic field dependence of (a) spin conductivity $\sigma$ and (b) thermal conductivity $\kappa$ are shown for several values of the anisotropy parameter $\Delta$. The magnetic field is measured in dimensionless units $H \equiv g\mu_B H/|J|$. The parameters used are $\alpha=0.01$, $|J_1|=|J|$, $|J_2|=0.8|J|$, and $g=2$ ( $S=1/2$).}
  \label{fig:Fig4}
  \end{figure*}

\subsection{Nonzero magnetic field}

In the presence of the magnetic field, the spin $\sigma(H)$ and thermal $\kappa(H)$ conductivities exhibit identical magnetic field dependence in the low-temperature regime, differing only by numerical prefactor, as detailed in Appendix~\ref{subsec:B2}. This behavior is illustrated in Fig.~\ref{fig:Fig4} for $T=10^{-4}$. The figure shows that increasing $H$ leads to a significant reduction in both $\sigma(H)$ and $\kappa(H)$. It follows that the magnetic field supresses the population of magnons.

In the Heisenberg model, the intrinsic gap is absent, and therefore the magnetic field is required to open an extrinsic gap in the system. The magnetic field provides a straightforward way to tune the magnitude of the gap. However, optimal transport properties require that the Zeeman gap satisfy the condition $\Delta^{\rm Zeeman}_{gp}\ll 1$, as only in this case we obtain the most efficient contributions to the transport quantities.

The WF law is independent of both the magnetic field $H$ and the anisotropy parameter $\Delta$.

\vspace{12mm}
\section{Summary and discussion}
In our work, we have theoretically investigated the spin and thermal transport properties of the $S=1/2$ ferromagnet on the four-sublattice FCC lattice (see Fig.~\ref{fig:Fig1}). Using LSWT and Kubo formalism, we have computed the temperature and magnetic field dependence of both the spin and thermal conductivity in the low-temperature regime for various values of the anisotropy parameter $\Delta$.

Our results complement the study of spin conductivity as a function of frequency \cite{sentef2007} and the analysis of transverse thermal transport \cite{pires2024} by investigating the temperature and magnetic-field dependence of longitudinal magnon transport on the FCC lattice.
  
Our analysis reveals that, in the absence of the magnon gap ($\Delta=1$ and $H=0$), both the spin conductivity $\sigma $ and the thermal conductivity $\kappa$ diverge due to the gapless magnon dispersion relation. Therefore, to make applications possible, the system must exhibit a gap in the dispersion relation. Moreover, this gap in the system must satisfy the condition $\Delta_{gp}\ll1$ to achieve the most efficient magnon transport. Despite the divergence of individual conductivities, their ratio remains finite and acquires the universal value $L=\kappa/\sigma T $, where $L=5/2$ in units of $(k_B/g\mu_B)^2$.  This shows excellent agreement with Ref.~\cite{nakata2017}, where this law was derived using the Boltzmann formalism for both the antiferromagnetic and ferromagnetic anisotropic Heisenberg models on a simple cubic lattice. This magnon WF law is independent of both the anisotropy parameter $\Delta$ and the magnetic field $H$, and depends only on the $g$-factor of the system. We also note that the dispersionless magnon branches do not contribute to either the spin conductivity or the thermal conductivity.

Although LSWT provides a valid description in the low-temperature regime by neglecting magnon-magnon interactions, its applicability breaks down at elevated temperatures as the number of magnons increases. To overcome this issue, higher-order terms in the HP transformation must be included. This inclusion of the interaction between magnons can be captured by allowing the Gilbert damping parameter to acquire temperature dependence. 

In the literature, it has been shown that the damping parameter for both three-dimensional and two-dimensional Heisenberg antiferromagnets in the classical limit scales with temperature as $\alpha(T)\propto T^2$ \cite{tyc1990,harris1971}.

Finally, our theoretical study establishes a framework for understanding and predicting the spin and thermal conductivity in FCC ferromagnets with $S=1/2$ in the low-temperature regime. This framework is directly relevant to materials exhibiting gapped magnon dispersion relations and small Gilbert damping parameter $\alpha$. These spin and thermal conductivities, obtained by the Kubo formalism, can be applied to other FCC lattices with arbitrary spin $S$ and small $\alpha$ by appropriately modifying the expressions for HP transformation. Thus, our work provides a theoretical foundation for future studies of magnon-mediated magneto-thermal transport properties in various magnetic lattices and contributes to the broader understanding of magnonics applications. Our work also highlights the critical roles of the exchange anisotropy parameter, magnetic field, and temperature in controlling the magnon transport properties of magnetic insulators.

\onecolumngrid
\appendix 
\section{Transport coefficients}
\label{sec:A}
In a three-dimensional magnon gas, the magnetization current (${\bm { \mathcal J}}^z_ M=\gamma {\bm {\mathcal J}}^z_{s}$, where $\gamma=g\mu_B/\hbar$) and thermal current are defined by \cite{aoya2019,aoya2022,rez}
\begin{equation}
{\bm { \mathcal  J}^z_M}=g\mu_B\sum_{\bm q}\sum_{\alpha=\pm}{ {\bm v^\alpha_{\bm q}}} b^\dagger_{\bm q\alpha}b_{\bm q\alpha}, \;\;\; \bm {\mathcal J}_{th}=\sum_{\bm q}\sum_{\alpha=\pm}\omega_{\bm q\alpha} \bm v^\alpha_{\bm q}b^\dagger_{\bm q\alpha}b_{\bm q\alpha},
\label{eq:A1}
\end{equation}
where 
\begin{equation}
\omega_{\bm q\pm}=\Delta^\pm_{gp}+D^\pm|\bm q|^2, \;\;\; \Delta^\pm_{gp}=|J_2|\Delta^\pm+g\mu_B H,
\label{eq:A2}
\end{equation}
\begin{equation}
v^\pm_{\bm q\mu}=2D^{\pm}q_{\mu},
\label{eq:A3}
\end{equation} 
and
\begin{equation}\Delta^\pm= \frac{\Delta(3+8\delta)-3}{2}\pm\frac{1}{2}\sqrt{C_1},\;\;\;D^{\pm}=\frac{|J_2|}{4}(1\mp\frac{C_2}{\sqrt{C_1}}),
\end{equation}
\begin{equation}
C_1=\left(\Delta(3+4\delta)-3\right)^2+48\delta^2, \;\;\; C_2=3+8\delta^2- \Delta(3+4\delta),
\label{eq:A4}	
\end{equation}
\begin{equation}
   \delta=|J_1|/|J_2|.
\end{equation}

In what follows, all transport quantities refer to their longitudinal components (e.g., $\sigma^{xx}=\sigma^{yy}=\sigma^{zz}$ while $\sigma^{\alpha \beta}=0$ for $\alpha \neq \beta$, with $\alpha, \beta =x,y,z$). The only distinction arises from the two dispersive magnon branches.

We present the general expressions for the spin and thermal conductivity in terms of the transport coefficients as follows:
\begin{equation}
\sigma^\pm(T)=L^\pm_{11},\;\;\; \kappa^\pm(T)=\frac{1}{T}\left(L^\pm_{22}-\frac{(L^\pm_{12})^2}{L^\pm_{11}} \right),
\label{eq:A5}
\end{equation}
where the transport coefficients in linear-response theory are given by 
\begin{equation}
L^\pm_{\alpha\beta}=-\lim_{\omega \to 0}\frac{Q^{\pm,R}_{\alpha\beta}(\omega)-Q^{\pm,R}_{\alpha\beta}(0)}{i\omega}.
\label{eq:A6}
\end{equation}
Writing the formula in a more compact form, we obtain 
\begin{equation}
L^{\pm}_{\alpha\beta}=i\frac{dQ^{\pm,R}_{\alpha\beta}(\omega)}{d\omega}\Big |_{\omega=0}
\label{eq:A7},
\end{equation}
where 
\begin{equation}
 Q^{\pm}_{11}(i\omega_n)=-\frac{1}{ \mathcal N}\int_{0}^{\beta}d\tau e^{i\omega_n \tau}\langle T_\tau { \mathcal J_{M,\pm}^z(\tau) \mathcal J^z_{M,\pm}(0)}\rangle,
 \label{eq:A8}
 \end{equation}
\begin{equation}
 Q^{\pm}_{12}(i\omega_n)=-\frac{1}{ \mathcal N}\int_{0}^{\beta}d\tau e^{i\omega_n \tau}\langle T_\tau { \mathcal J_{M,\pm}^z(\tau) \mathcal J_{th,\pm}(0)}\rangle,
 \label{eq:A9}
 \end{equation}
\begin{equation}
 Q^{\pm}_{22}(i\omega_n)=-\frac{1}{ \mathcal N}\int_{0}^{\beta}d\tau e^{i\omega_n \tau}\langle T_\tau { \mathcal J_{th,\pm}(\tau) \mathcal J_{th,\pm}(0)}\rangle.
 \label{eq:A10}
 \end{equation}
 and $\omega_n=2\pi n/\beta$ is the bosonic Matsubara frequency.
 
We define Green's function for a free magnon gas 
\begin{equation}
 \mathcal D_{\bm q\pm}(\tau)=-\langle T_\tau b_{\bm q\pm}(\tau)b^\dagger_{\bm q\pm}(0)\rangle=\frac{1}{\beta}\sum_{i\omega_m}\mathcal D_{\bm q\pm}(i\omega_m)e^{-i\omega_m\tau}.
 \label{eq:A11}
\end{equation}

Inserting Eq.~(\ref{eq:A1}) into Eqs.~(\ref{eq:A8})--(\ref{eq:A10}), we obtain \cite{aoya2019,aoya2022,abrikosov}
\begin{equation}
\frac{1}{\beta}\sum_{i\omega_m}\mathcal D_{\bm q\pm}(i\omega_m)\mathcal D_{\bm q\pm}(i\omega_n+i\omega_m)=\int_{-\infty}^{\infty}\frac{dx}{2\pi i}\left(\mathcal D^R_{\bm q\pm}(x)-\mathcal D^A_{\bm q\pm}(x)\right)\left(\mathcal D^R_{\bm q\pm}(x+i\omega_n)+\mathcal D^A_{\bm q\pm}(x-i\omega_n)\right)n_B(x),
\label{eq:A12}
\end{equation}
where 
\begin{equation}
\mathcal D^R_{\bm q\pm}(\omega)=\frac{1}{\omega-\omega_{\bm q\pm}+i\alpha \omega}.
\label{eq:A13}
\end{equation}
Magnon Green's function explicitly include the Gilbert damping parameter $\alpha$. This parameter plays a critical role in ensuring that derived physical quantities remain finite.

By analytically continuing the Matsubara frequency $i\omega_n \to \omega+i0^+$ and then taking the derivative with respect to $\omega$ as indicated in Eq.~(\ref{eq:A7}), we obtain 

\begin{equation}
L^{\pm}_{\alpha\beta}=\frac{1}{4\pi}\int_{-\infty}^{\infty}dx \frac{dn_B(x)}{dx}\mathcal L^\pm_{\alpha\beta}(x),
\label{eq:A14}
\end{equation}
where

\begin{equation*}
\mathcal L^\pm_{11}=\frac{(g\mu_B)^2}{\mathcal N}\sum_{\bm q}(v^\pm_{\bm q \mu })^2(\mathcal D^R_{\bm q\pm}(x)-\mathcal D^A_{\bm q\pm}(x))^2=-(g\mu_B)^2K_{-3/2}\int_{0}^{\infty}\frac{u^{3/2} du}{\left[\left(\frac{x-\Delta_{gp}^\pm}{k_BT}-u\right)^2+\Big(\frac{\alpha x}
{k_BT}\Big)^2\right]^2}
\end{equation*}
\begin{equation}
=-\frac{2}{3\pi}\frac{  (g\mu_B)^2}{\alpha  \sqrt{D^\pm}}\theta(x-\Delta^\pm_{gp})\frac{(x-\Delta^\pm_{gp})^{3/2}}{x},
\label{eq:A15}
\end{equation}
\begin{equation*}
\mathcal L^\pm_{12}=\frac{g\mu_B}{\mathcal N}\sum_{\bm q}\omega_{\bm q\pm}(v^\pm_{\bm q\mu })^2(\mathcal D^R_{\bm q\pm}(x)-\mathcal D^A_{\bm q\pm}(x))^2=\frac{\Delta^\pm_{gp}}{g\mu_B}\mathcal L^\pm_{11}-g\mu_BK_{-1/2}\int_{0}^{\infty}\frac{u^{5/2} du}{\left[\left(\frac{x-\Delta^\pm_{gp}}{k_BT}-u\right)^2+\Big(\frac{\alpha x}
{k_BT}\Big)^2\right]^2}
\end{equation*}
\begin{equation}
=\frac{\Delta^\pm_{gp}}{g\mu_B}\mathcal L^\pm_{11}-\frac{2}{3\pi}\frac{ g\mu_B}{\alpha  \sqrt{D^\pm}}\theta(x-\Delta^\pm_{gp})\frac{(x-\Delta^\pm_{gp})^{5/2}}{x},
\label{eq:A16}
\end{equation}
\begin{equation*}
\mathcal L^\pm_{22}=\frac{1}{\mathcal N}\sum_{\bm q}(\omega_{\bm q\pm})^2(v^\pm_{\bm q\mu })^2(\mathcal D^R_{\bm q\pm}(x)-\mathcal D^A_{\bm q\pm}(x))^2=-\left(\frac{\Delta_{gp}^\pm}{g\mu_B}\right)^2\mathcal L^\pm_{11}+2\frac{\Delta^\pm_{gp}}{g\mu_B}\mathcal L_{12}^\pm-K_{1/2}\int_{0}^{\infty}\frac{u^{7/2} du}{\left[\left(\frac{x-\Delta^\pm_{gp}}{k_BT}-u\right)^2+\Big(\frac{\alpha x}
{k_BT}\Big)^2\right]^2}
\end{equation*}
\begin{equation}
=-\left(\frac{\Delta^\pm_{gp}}{g\mu_B}\right)^2 \mathcal L^\pm_{11}+2\frac{\Delta^\pm_{gp}}{g\mu_B}\mathcal L_{12}^\pm-\frac{2}{3\pi}\frac{1}{\alpha  \sqrt{D^\pm}}\theta(x-\Delta^\pm_{gp})\frac{(x-\Delta^\pm_{gp})^{7/2}}{x},
\label{eq:A17}
\end{equation}
where $K_\beta$ is given by
\begin{equation}
K_\beta=\frac{4}{3\pi^2}\frac{\alpha^2x^2}{\sqrt{D^\pm}}(k_BT)^{\beta}.
\label{eq:A18}
\end{equation}
In the above formulas, we have used the limit $\mathcal N \to \infty$, replacing the sum with the integral:
\begin{equation}
\frac{1}{\mathcal N}\sum_{\bm q} \to\frac{1}{(2\pi)^3}\int q^2dq\int \sin\theta d\theta\int d\phi.
\label{eq:A19}
\end{equation}
We have used the following identity for $\alpha \ll1$
\begin{equation}
\frac{1}{[(x-u)^2+(\alpha x)^2]^2}=\frac{\pi}{2(\alpha x)^3}\delta(x-u).
\label{eq:A20}
\end{equation}


By inserting Eqs.~(\ref{eq:A15})--(\ref{eq:A17}) into Eq.~(\ref{eq:A14}), we obtain integrals for both diagonal and mixed transport coefficients
\begin{equation}
L^\pm_{11}=-\frac{ (g\mu_B)^2}{6\pi^2\alpha  (D^\pm)^{1/2}}\int_{\Delta^\pm_{gp}}^{\infty}dx \frac{dn_B(x)}{dx}\frac{(x-\Delta^\pm_{gp})^{3/2}}{x},
\label{eq:A21}
\end{equation}
\begin{equation}
L^\pm_{12}=\frac{\Delta^\pm_{gp}}{g\mu_B} L^\pm_{11}-\frac{ g\mu_B}{6\pi^2\alpha  (D^\pm)^{1/2}}\int_{\Delta^\pm_{gp}}^{\infty}dx \frac{dn_B(x)}{dx}\frac{(x-\Delta^\pm_{gp})^{5/2}}{x},
\label{eq:A22}
\end{equation}
\begin{equation}
L^\pm_{22}=-\left(\frac{\Delta^\pm_{gp}}{g\mu_B}\right)^2 L^\pm_{11}+2\frac{\Delta^\pm_{gp}}{g\mu_B} L^\pm_{12}-\frac{1}{6\pi^2\alpha  (D^\pm)^{1/2}}\int_{\Delta^\pm_{gp}}^{\infty}dx \frac{dn_B(x)}{dx}\frac{(x-\Delta^\pm_{gp})^{7/2}}{x}.
\label{eq:A23}
\end{equation}
To proceed, we introduce the integral 
\begin{equation}
\int_{\Delta}^{\infty}dx\frac{dn_B(x)}{dx} \frac{(x-\Delta)^{\mu/2}}{x}=
\begin{cases}
	-\Gamma(\mu/2+1)(k_BT)^{\mu/2}e^{-\Delta/k_BT}/\Delta, & \text{when } \Delta/k_BT \gg 1 \\\\
-k_BT \Delta^{\mu/2-2}B(\mu/2+1,2-\mu/2), & \text{when } \Delta/k_BT \ll 1
\end{cases}
\label{eq:A24}
\end{equation}
where $B(z_1,z_2)=\Gamma(z_1)\Gamma(z_2)/\Gamma(z_1+z_2)$ is Beta function.

\section{Spin conductivity, thermal conductivity, and the Wiedemann-Franz law }
\label{sec:B}
\subsection{Zero magnetic field}
\label{subsec:B1}
 Using the analytical expressions for the transport coefficients derived in the previous section, we obtain the temperature dependence of $\sigma(T)$ and $\kappa(T)$ by summing the contributions from both dispersive magnon branches 
\begin{equation}
\sigma(T)/(g\mu_B)^2=\frac{1}{8\alpha \pi^{3/2}}\left(\frac{k_B T}{|J_2|} \right)^{3/2}\sum_{\gamma=\pm}\sqrt{\frac{|J_2|}{D^\gamma}}\frac{1}{\Delta^\gamma} e^{-|J_2|\Delta^\gamma/k_BT},
\label{eq:B1}
\end{equation}
\begin{equation}
\kappa(T)/k_B|J_2|=\frac{5}{16\alpha\pi^{3/2}}\left( \frac{k_BT}{|J_2|}\right)^{5/2}\sum_{\gamma=\pm}\sqrt{\frac{|J_2|}{D^\gamma}}\frac{1}{\Delta^\gamma}e^{-|J_2|\Delta^\gamma/k_BT}.
\label{eq:B2}
\end{equation}

Wiedemann-Franz law is defined as
\begin{equation}
\frac{\kappa^\pm}{\sigma^\pm T}=\frac{1}{T^2}\left[\frac{L^\pm_{22}}{L^\pm_{11}}-\left(\frac{L^\pm_{12}}{L^\pm_{11}}\right)^2\right]=L,
\label{eq:B3}
\end{equation}
where 
\begin{equation}
L=\left[\frac{\Gamma(9/2)}{\Gamma(5/2)}-\left(\frac{\Gamma(7/2)}{\Gamma(5/2)}\right)^2 \right]\left(\frac{k_B}{g\mu_B}\right)^2=\frac{5}{2}\left(\frac{k_B}{g\mu_B} \right)^2.
\label{eq:B4}
\end{equation}

\subsection{Nonzero magnetic field}
\label{subsec:B2}
In the presence of the nonzero magnetic field ($H\neq0$), both the spin and thermal conductivities develop explicit dependence on $H$

 \begin{equation}
\sigma(H)/(g\mu_B)^2=\frac{1}{8\alpha \pi^{3/2}}\left(\frac{k_B T}{|J_2|} \right)^{3/2}\sum_{\gamma=\pm}\sqrt{\frac{|J_2|}{D^\gamma}}\frac{1}{(\Delta^\gamma+g\mu_B H/|J_2|)} e^{-|J_2|\Delta^\gamma/k_BT}e^{-\frac{g\mu_BH}{|J_2|} \frac{|J_2|}{k_BT}},
\label{eq:B6}
\end{equation}
\begin{equation}
\kappa(H)/k_B|J_2|=\frac{5}{16\alpha\pi^{3/2}}\left( \frac{k_BT}{|J_2|}\right)^{5/2}\sum_{\gamma=\pm}\sqrt{\frac{|J_2|}{D^\gamma}}\frac{1}{(\Delta^\gamma+g\mu_BH/|J_2|)}e^{-|J_2|\Delta^\gamma/k_BT}e^{-\frac{g\mu_BH}{|J_2|} \frac{|J_2|}{k_BT}}.
\label{eq:B7}
\end{equation}

\twocolumngrid
\bibliography{magnon_transport}

\end{document}